\documentclass[aps,pra,preprintnumbers,showpacs,tightenlines]{revtex4}

\usepackage{amssymb}
\usepackage{amsmath}
\usepackage{graphicx}
\usepackage{epsfig}
\usepackage{subfigure}
\usepackage{amsfonts}
\usepackage{CJK}

\begin{document}

\title{Generating multipartite entangled states of qubits distributed in different cavities}

\author{Xiao-Ling He$^{1}$, Qi-Ping Su$^{2}$, Feng-Yang Zhang$^{3}$, and Chui-Ping Yang$^{2}$}
\email{yangcp@hznu.edu.cn}

\address{$^1$School of Science, Zhejiang University of Science and Technology, Hangzhou, Zhejiang 310023, China}

\address{$^2$Department of Physics, Hangzhou Normal University, Hangzhou, Zhejiang 310036, China}

\address{$^3$School of Physics and Materials Engineering, Dalian Nationalities University, Dalian 116600, China}

\date{\today}

\begin{abstract}
Cavity-based large-scale quantum information processing (QIP) needs a large number of qubits and placing all of them in a single cavity quickly runs into many fundamental and practical problems such as the increase of cavity decay rate and decrease of qubit-cavity coupling strength. Therefore, future QIP most likely will require quantum networks consisting of a large number of cavities, each hosting and coupled to multiple qubits. In this work, we propose a way to prepare a $W$-class entangled state of spatially-separated multiple qubits in different cavities, which are connected to a coupler qubit. Because no cavity photon is excited, decoherence caused by the cavity decay is greatly suppressed during the entanglement preparation. This proposal needs only one coupler qubit and one operational step, and does not require using a classical pulse, so that the engineering complexity is much reduced and the operation is greatly simplified. As an example of the experimental implementation, we further give a numerical analysis, which shows that high-fidelity generation of the $W$ state using three superconducting phase qubits each embedded in a one-dimensional transmission line resonator is feasible within the present circuit QED technique. The proposal is quite general and can be applied to accomplish the
same task with other types of qubits such as superconducting flux qubits, charge qubits, quantum dots, nitrogen-vacancy centers and atoms.
\end{abstract}

 \maketitle
\date{\today}

\section{INTRODUCTION}

Entanglement is a key resource of quantum information processing (QIP) and
quantum communication. During the past decade, a large number of proposals
have been presented for entanglement generation. Although most of the
quantum information protocols focus on bipartite systems, multipartite
entanglement has also attracted much interest because of its potential
applications in QIP and quantum communication. It has been shown [1] that
there exist two inequivalent classes of multipartite entangled states, i.e.,
Greenberger-Horne-Zeilinger (GHZ) states [2] and $W$ states [1], which can
not be converted to each other by local operations and classical
communications. With respect to the tripartite entangled states, it was
shown [1] that $W$ states are robust against losses of qubits since they
retain bipartite entanglement if we trace out any one qubit, whereas GHZ
states are fragile since the remaining bipartite states are
separable states. This property turns $W$ states very attractive for various
quantum communication tasks. For instances, the $W$ states can be used as
quantum channels for teleportation of entangled pairs [3], quantum
teleportation [4], quantum key distribution [5] and so on. During the past years,
many theoretical schemes for generating $W$ states have been proposed. For examples,
(i) schemes have been proposed to generate $W$ states in trapped
ions [6,7], atomic ensembles [8], Ising chains with nearest-neighbor coupling
by global control [9], or photons on-chip multiport photonic lattices [10];
(ii) by using linear optical elements and photon detection, schemes
have been proposed to generate $W$ states of spatially-separated distant
atoms [11] or photons [12]; (iii) by using parametric down conversion, schemes
have been presented to generate $W$ states of photons [13]; and (iv)
based on cavity QED, how to prepare $W$ states has been proposed in quantum dots
coupled to a cavity [14], superconducting qubits embedded within a single
cavity [15,16], or atoms interacting with a cavity [17,18]. On the other hand,
the $W$ states have been experimentally created with up to eight
trapped ions [19], four optical modes [20], three superconducting phase
qubits coupled capacitively [21], and atomic ensembles in four quantum
memories [22], as well as two superconducting phase qubits plus a resonant
cavity [23].

The physical system, composed of cavities and qubits, has attracted much attention for QIP.
Over the past twenty years, a large number of theoretical and experimental works have been done for implementing quantum information transfer, quantum logical gates, and quantum entanglement with qubits placed inside a {\it single} cavity or coupled to a resonator. These works are important in QIP based on cavity QED. However, they are valid only for the case that all qubits are placed in the same cavity or
coupled to a common resonator.

Attention is now shifting to large-scale QIP based on cavity QED, which needs a large number of qubits. Note that placing all of qubits in a single cavity quickly runs into many fundamental and practical problems such as the increase of cavity decay rate and decrease of qubit-cavity coupling strength. Therefore, future cavity-based QIP most likely will require quantum networks consisting of a large number of cavities, each hosting and coupled to multiple qubits. In this type of architecture, transfer and exchange of quantum information will not only occur among qubits in the same cavity but also happens between different cavities. Hence, attention must be paid to the preparation of quantum states of two or more cavities, preparation of quantum states of qubits located in different cavities, and implementation of quantum logic gates on qubits distributed over different cavities in a network. All of these ingredients are essential to realizing large-scale QIP based on cavity QED.

Motivated by the above, in this work we focus on how to prepare $W$ states of qubits distributed in many different cavities. Besides its use in
large-scale QIP, this work may be also interesting from the following point of view:

The prepared $W$ state can be stored in matter qubits with long
decoherence time. Once the $W$ state is needed for quantum communication,
one can transfer the $W$ state of matter qubits onto cavity photons and then
transmit the cavity photons to distant spatially-separated users located at
different nodes in a network. This can be achieved as follows. First, by local operations
within every cavity (i.e., a local operation is performed on a qubit and a cavity
in which the qubit is placed, so that the state of the qubit is transferred onto
the cavity photon), one can transfer the $W$ state of matter qubits onto the
cavity photons. Second, to transmit a cavity photon to a distant user in a network,
one can increase the cavity-decay rate (e.g., by adjusting
the mirrors at the end of an optical cavity or lowering the cavity quality
factor for a circuit cavity) to have the cavity photon leaked into an optical fiber,
which connects the cavity with the distant user. In this way, the $W$ state of the cavity
photons can be shared by different users in a network, and can be used
as a quantum channel for carrying out quantum communication tasks.

In the following, we will present a way for preparing $W$ states of qubits
distributed in $n$ different cavities. As shown below, this proposal has the following advantages:
(i) the entanglement preparation is performed
without excitation of the cavity photons, and thus decoherence induced by
the cavity decay is greatly suppressed; (ii) only one
coupler qubit is needed, one operational step is required, and no classical
pulse is used, hence the engineering complex is much reduced and the operation is
greatly simplified; and (iii) the operation time decreases as the number of
qubits increases.

This proposal is quite general, and can be applied to accomplish the
same task with different types of qubits, such as quantum dots, atoms, NV centers,
superconducting qubits (e.g., phase, flux and charge qubits), and so on. To the best of our knowledge,
how to create the $W$ state of qubits, distributed in different cavities connecting
to a coupler qubit, has not been reported so far.

This paper is organized as follows. In Sec. 2, we show how to generate the $%
W$ state of $n$ qubits distributed in $n$ different cavities. In Sec. 3,
as an example, we analyze the experimental feasibility of preparing the $W$
state of three superconducting phase qubits, which are distributed in three
different one-dimensional transmission line resonators. A concluding summary
is enclosed in Sec.~4.

\section{\textit{W-}STATE PREPARATION}

In this section, we first construct a Hamiltonian for the $W$ state
preparation. We then give a discussion on how to prepare the $W$ state of $n$
qubits $\left( 1,2,...,n\right) $ distributed in the $n$ cavities.

\subsection{Hamiltonian}

\begin{figure}[tbp]
\begin{center}
\includegraphics[bb=83 159 525 703, width=10.0 cm, clip]{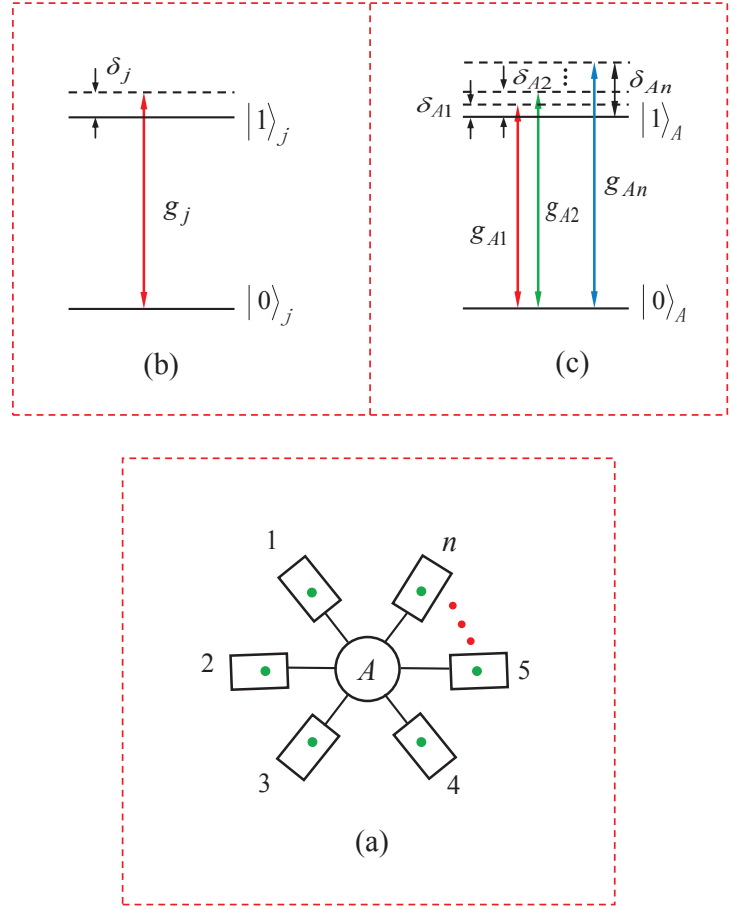} %
\vspace*{-0.08in}
\end{center}
\caption{(color online) (a) Diagram of a coupler qubit $A$ (a circle at the
center) and $n$ cavities each hosting a qubit. A dark
square represents a cavity while a green dot labels a qubit placed in each
cavity, which can be an atom or a solid-state qubit. The coupler qubit $A$
can be an atom or a quantum dot, and can also be a superconducting qubit
capacitively or inductively coupled to each cavity. (b) Cavity $j$
dispersively coupled to qubit $j$ (placed in cavity $j$) with coupling constant $g_j$ and
detuning $\delta_j$. (c) The coupler qubit $A$
dispersively interacting with $n$ cavities simultaneously, with
coupling constant $g_{Aj}$ and detuning $\delta_{Aj}$ for cavity $j$ ($j=1,2,...,n$).
Here, $\delta_{Aj}=\delta_j$, which holds for identical qubits $A$ and $j$.
Note that in (a), only one qubit in each cavity is drawn for simplicity.
In reality, for a quantum processor with multiple registers---each register consists of
a cavity and qubits in the cavity, more than one qubit are usually placed in each cavity.
To prepare the $W$ state of $n$ qubits each in a different cavity, only one qubit in each cavity is involved in the entanglement
preparation, while other qubits in each cavity can be made to be decoupled from their cavity
by adjusting their level spacings (e.g., solid-state qubits) or by
moving them out of their cavity (e.g., atomic qubits), such that they do not participate
during the $W$ state preparation.}
\label{fig:1}
\end{figure}

Consider $n$ cavities $(1,2,...,n)$ connected to a coupler qubit $A$, as
illustrated in Fig.~1(a). Cavity $j$ ($j=1,2,...n$) hosts qubit $j$, shown
as a black dot. Each qubit here has two levels $\left\vert 0\right\rangle $
and $\left\vert 1\right\rangle .$ Assume that the coupling constant of qubit
$j$ with cavity $j$ is $g_{j}.$ The coupler qubit $A$ in Fig.~1 interacts
with $n$ cavities $(1,2,...,n)$ simultaneously. We denote $g_{Aj}$ as the
coupling constant of qubit $A$ with cavity $j.$ In the interaction picture
under the free Hamiltonian of the whole system and applying the
rotating-wave approximation, we have
\begin{equation}
H_{I}=\sum_{j=1}^{n}g_{j}\left( e^{i\delta _{j}t}a_{j}\sigma
_{j}^{+}+h.c.\right) +\sum_{j=1}^{n}g_{Aj}\left( e^{i\delta
_{Aj}t}a_{j}\sigma _{A}^{+}+h.c.\right) ,
\end{equation}
where $\sigma _{j}^{+}=\left\vert 1\right\rangle _{j}\left\langle
0\right\vert $ and $\sigma _{A}^{+}=\left\vert 1\right\rangle
_{A}\left\langle 0\right\vert $ are, respectively, the raising operators for
qubit $j$ and qubit $A$, $\delta _{j}=\omega _{10j}-\omega _{cj}$ is the
detuning of the transition frequency $\omega _{10j}$ of qubit $j$ from the
frequency $\omega _{cj}$ of cavity $j,$ $\delta _{Aj}=\omega _{10A}-\omega
_{cj}$ is the detuning of the transition frequency $\omega _{10A}$ of qubit $%
A$ from the frequency $\omega _{cj}$ of cavity $j$ [Fig.~1(b,c)], and $a_{j}$
is the annihilation operator for the mode of cavity $j$ ($j=1,2,...,n$).

In the case $\delta _j\gg g_j$ and $\delta _{Aj}\gg g_{Aj},$ there is no
energy exchange between the qubit system and the cavities. In addition,
under the condition of
\begin{equation}
\frac{\left\vert \delta _{A(j+1)}-\delta _{Aj}\right\vert }{\delta
_{Aj}^{-1}+\delta _{A(j+1)}^{-1}}\gg g_{Aj}g_{A(j+1)},
\end{equation}
there is no interaction between the $n$ cavities, which is induced by the
coupler qubit $A.$ Hence, we can obtain [24,25]
\begin{eqnarray}
H_{\mathrm{eff}} &=&-\sum_{j=1}^{n}\frac{g_{j}^{2}}{\delta _{j}}\left(
\left\vert 0\right\rangle _{j}\left\langle 0\right\vert
a_{j}^{+}a_{j}-\left\vert 1\right\rangle _{j}\left\langle 1\right\vert
a_{j}a_{j}^{+}\right)  \nonumber \\
&&-\ \sum_{j=1}^{n}\frac{g_{Aj}^{2}}{\delta _{Aj}}\left( \left\vert
0\right\rangle _{A}\left\langle 0\right\vert a_{j}^{+}a_{j}-\left\vert
1\right\rangle _{A}\left\langle 1\right\vert a_{j}a_{j}^{+}\right)  \nonumber
\\
&&+\sum_{j=1}^{n}\lambda _{j}\left[ e^{i(\delta _{j}-\delta _{Aj})t}\sigma
_{j}^{+}\sigma _{A}+h.c.\right]
\end{eqnarray}
where $\lambda _{j}=\frac{g_{j}g_{Aj}}{2}\left( 1/\delta _{j}+1/\delta
_{Aj}\right) .$ The first (second) term of Eq. (3) describes the
photon-number dependent Stark shifts of qubit $j$ (qubit $A$), while the
last term describes the \textquotedblleft dipole\textquotedblright\ coupling
between qubit $j$ and qubit $A$ mediated by the mode of cavity $j$.

Assume that each cavity is initially in the vacuum state, and set
\begin{equation}
\delta _{j}=\delta _{Aj}.
\end{equation}
Then the Hamiltonian (3) reduces to
\begin{equation}
H_{\mathrm{eff}}=H_{0}+H_{\mathrm{int}},
\end{equation}
with
\begin{eqnarray}
H_{0} &=&\sum_{j=1}^{n}\frac{g_{j}^{2}}{\delta _{j}}\left\vert
1\right\rangle _{j}\left\langle 1\right\vert +\sum_{j=1}^{n}\frac{g_{Aj}^{2}%
}{\delta _{Aj}}\left\vert 1\right\rangle _{A}\left\langle 1\right\vert , \\
H_{\mathrm{int}} &=&\sum_{j=1}^{n}\lambda _{j}\left( \sigma _{j}^{+}\sigma
_{A}^{-}+\sigma _{j}^{-}\sigma _{A}^{+}\right) .
\end{eqnarray}
Note that the Hamiltonians (6) and (7) do not contain the operators of the
cavity fields. Thus, only the state of the qubit system undergoes an
evolution under the Hamiltonians (6) and (7). Therefore, each cavity field
is virtually excited.

In a new interaction picture under the Hamiltonian $H_{0}$ and using the
following condition
\begin{eqnarray}
\frac{g_{1}^{2}}{\delta _{1}} &=&\frac{g_{2}^{2}}{\delta _{2}}=\cdot \cdot
\cdot =\frac{g_{n}^{2}}{\delta _{n}}=\chi , \\
\frac{g_{k}^{2}}{\delta _{k}} &=&\sum_{j=1}^{n}\frac{g_{Aj}^{2}}{\delta _{Aj}%
},\;k\in \{1,2,...,n\}
\end{eqnarray}
we can obtain
\begin{equation}
\widetilde{H}_{\mathrm{int}}=e^{iH_0t}H_{\mathrm{int}}e^{-iH_0t}=H_{\mathrm{%
int}}.
\end{equation}
In addition, we set
\begin{equation}
\frac{g_1g_{A1}}{\delta _1}=\frac{g_2g_{A2}}{\delta _2}=...=\frac{g_ng_{An}}{%
\delta _n}=\lambda ,
\end{equation}
which is equivalent to $\lambda _1=\lambda _2=...=\lambda _n=\lambda $ under
the condition (4) and because of the $\lambda _j$'s expression listed below Eq. (3). Thus, we can express the Hamiltonian (10) as
\begin{equation}
\widetilde{H}_{\mathrm{int}}=\lambda \left( J_{+}\sigma _A^{-}+J_{-}\sigma
_A^{+}\right) ,
\end{equation}
where $J_{+}=\sum_{j=1}^n\sigma _j^{+}$ and $J_{-}=\sum_{j=1}^n\sigma
_j^{-}. $ This constructed Hamiltonian (12) will be employed for preparing
the $n$ intracavity qubits ($1,2,...,n$) in the $W$ state, as shown below.

As most related to this work, we should mention a Hamiltonian of $%
J_{+}a+J_{-}a^{+}.$ As is well known, this Hamiltonian can be used to create
an $n$-qubit $W$ state. However, this Hamiltonian is for a system composed
of $n$ qubits ($1,2,...,n$) \textit{simultaneously} interacting with a single
{\it common} cavity, described by a photon creation operator $a^{+}$
and annihilation operator $a$. Thus, the system characterized by the
Hamiltonian $J_{+}a+J_{-}a^{+}$ is different from our current one, i.e., a
system consisting of $n$ qubits interacting with $n$ different cavities.
Furthermore, both systems are quite different in the qubit-cavity
coupling mechanism. Finally, as discussed in the introduction, this work is
based on different motivations.

The present work differs from the one in Ref. [9]. The latter discussed
how to prepare a $W$ state of multiple qubits based on a one-dimensional Ising chain with
\textit{nearest-neighbor} coupling by a global control. One can see that our
Hamiltonian (12) constructed above does not contain a term $\sigma _{\alpha
,j}\sigma _{\beta ,j+1}+h.c.$ describing the nearest-neighbor coupling.
Here, $\sigma _{\alpha ,j}$ and $\sigma _{\beta ,j+1}$ are the Pauli
operators of the qubits $j$ and $j+1$, respectively ($\alpha ,\beta \in
\{x,y,z\}$).

\subsection{$W$\textbf{-state preparation}}

Let us assume that: (i) each cavity is initially in the vacuum state; (ii)
each intracavity qubit is initially in the ground state, i.e., qubit $j$ is
in the state $\left\vert 0\right\rangle _{j}$, and all intracavity qubits
are decoupled from their respective cavities; and (iii) the coupler qubit $A$
is initially in the state $\left\vert 1\right\rangle _{A}$ and decoupled
from the $n$ cavities. The decoupling of each qubit from its cavity
(cavities) can be achieved by prior adjustment of the qubit's level
spacings. For superconducting devices, their level spacings can be rapidly
adjusted by varying external control parameters (e.g., magnetic flux applied
to phase, transmon, or flux qutrits; see, e.g., [26-28]).

To generate the $W$ state, we now adjust the level spacings of all qubits
(including the coupler qubit $A$) to have the state of the qubit system
undergo the time evolution described by the Hamiltonian (12). Based on the
Hamiltonian (12) and after returning to the original interaction picture by
performing a unitary transformation $e^{-iH_{0}t},$ it is easy to find that
the initial state $\prod\limits_{j=1}^{n}\left\vert 0\right\rangle
_{j}\left\vert 1\right\rangle _{A}$ of the qubit system evolves into
\begin{equation}
e^{-i\chi t}\left[ \cos \left( \sqrt{n}\lambda t\right)
\prod\limits_{j=1}^{n}\left\vert 0\right\rangle _{j}\otimes \left\vert
1\right\rangle _{A}-i\sin \left( \sqrt{n}\lambda t\right) \left\vert
W_{n-1,1}\right\rangle \otimes \left\vert 0\right\rangle _{A}\right] ,
\end{equation}
where the term in brackets was obtained under the Hamiltonian (12) while the
factor $e^{-i\chi t}$ was achieved by performing the unitary transformation $%
e^{-iH_{0}t}$ and using Eqs. (8) and (9). Here, the state $\left\vert
W_{n-1,1}\right\rangle $ of the $n$ qubits $(1,2,...,n)$ is given by
\begin{equation}
\left\vert W_{n-1,1}\right\rangle =\frac{1}{\sqrt{n}}\sum P_{z}\left\vert
0\right\rangle ^{\otimes \left( n-1\right) }\left\vert 1\right\rangle ,
\end{equation}
where $P_{z}$ is the symmetry permutation operator for the qubits $(1,2,...,n),$
and $\sum P_{z}\left\vert 0\right\rangle ^{\otimes \left( n-1\right) }\left\vert
1\right\rangle $ denotes the totally symmetric state in which $n-1$ of
qubits $(1,2,...,n)$ are in the state $\left\vert 0\right\rangle $ while the
remaining qubit is in the state $\left\vert 1\right\rangle .$ For instance,
we have $\left\vert W_{2,1}\right\rangle =\frac{1}{\sqrt{3}}\left(
\left\vert 001\right\rangle +\left\vert 010\right\rangle +\left\vert
100\right\rangle \right) $ when $n=3.$ The state (14) is known as the $W$%
-class entangled state in the context of quantum information [1]. From Eq.
(13), one can see that the $W$ state (14) of qubits $(1,2,...,n)$ can be
created when the interaction time equals to $t=\pi /\left( 2\sqrt{n}\lambda
\right)$, which decreases as the number $n$ of qubits increases.

To freeze the prepared $W$ state, the level spacings for each
qubit need to be adjusted back to the original configuration, such that each
qubit is decoupled from the cavities.

We should mention that adjusting the qubit level spacings is unnecessary.
Alternatively, the coupling or decoupling of the qubits with the cavities
can be obtained by adjusting the frequency of each cavity. The rapid tuning
of cavity frequencies has been demonstrated in superconducting microwave
cavities (e.g., in less than a few nanoseconds for a superconducting
transmission line resonator [29]).

\subsection{Discussion}

Let us now discuss the issues which are most relevant to the experimental
implementation of the method. For the method to work, the following
requirements need to be satisfied:

(i) The conditions (2), (4), (8) and (9) need to be met. The condition (2)
can be reached by prior adjustment of the frequency of each cavity. The
condition (4) is automatically ensured for the identical qubits. Given\ $%
\delta _{1},\delta _{2},...,$and $\delta _{n},$ the condition (8) can be met
by adjusting the coupling constants $g_{1},g_{2},...,$ and $g_{n}$ (e.g.,
for solid-state qubits, the qubit-cavity coupling constants can be readily
changed by varying the positions of the qubits embedded in their cavities).
The condition (9) can be met by setting
\begin{equation}
g_{Aj}/g_{j}=1/\sqrt{n},
\end{equation}
where $j=1,2,...,n$. Given\ $g_{j}$, this requirement (15) can be obtained
by adjusting $g_{Aj}$ (e.g., for a solid-state coupler qubit $A$, $g_{Aj}$
can be adjusted by changing the qubit-cavity coupler capacitance $C_{j},$
see Fig.~2).

(ii) The operation time required for the entanglement preparation
needs to be much shorter than the energy relaxation time $T_{1}$
and dephasing time $T_{2}$ of the level $\left\vert 1\right\rangle $, such
that the decoherence, caused by energy relaxation and dephasing of the
qubits, is negligible during the operation.

(iii) For cavity $i$ ($i=1,2,...,n$), the lifetime of the cavity mode is
given by $T_{\mathrm{cav}}^{i}=\left( Q_{i}/2\pi \nu _{c,i}\right) /%
\overline{n}_{i},$ where $Q_{i}$ and $\overline{n}_{i}$ are the (loaded)
quality factor and the average photon number of cavity $i$, respectively.
For the $W$-state preparation, the lifetime of the cavity modes is given by
\begin{equation}
T_{\mathrm{cav}}=\frac{1}{n}\min \{T_{\mathrm{cav}}^{1},T_{\mathrm{cav}%
}^{2},...,T_{\mathrm{cav}}^{n}\},
\end{equation}
which should be much longer than the operation time$,$ such that the effect
of cavity decay is negligible for the operation.

(iv) When the coupler qubit $A$ is a solid-state qubit, there may exist an
intercavity cross coupling during the operation, which should be negligibly
small. As an example, let us consider that each cavity is coupled to qubit $%
A $ through a coupler capacitance. In this case, the intercavity cross
coupling is mostly determined by the coupling capacitances $%
C_{1},C_{2},...,C_{n}$ and the qutrit's self capacitance $C_{q}$, because
the field leakage through space is extremely low for high-$Q$\ resonators as
long as the inter-cavity distance is much greater than the transverse
dimension of the cavities. As our numerical simulations, shown by Fig.~4
below, the effects of the inter-cavity coupling can however be made
negligible as long as $g_{kl}\leq 0.2g_{\max }$ with $g_{\max }=\max
\{g_{A1},g_{A2},...,g_{An}\}$, where $g_{kl}$ is the corresponding
intercavity coupling constant between any two cavities $k$ and $l$ of the $n$
cavities ($1,2,...,n$).

\begin{figure}[tbp]
\begin{center}
\includegraphics[bb=176 331 433 558, width=7.5 cm, clip]{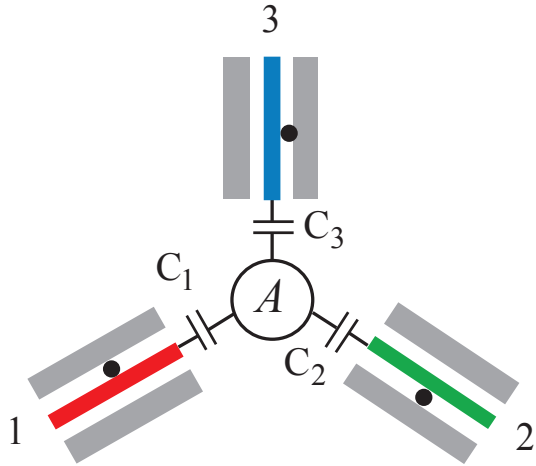} %
\vspace*{-0.08in}
\end{center}
\caption{(color online) Setup for three cavities (1,2,3) coupled by a
superconducting phase qubit $A$. Each cavity here is a one-dimensional
coplanar waveguide transmission line resonator. The circle $A$ represents a
superconducting phase qubit, which is capacitively coupled to cavity j via a
capacitance $Cj$ ($j = 1,2,3$). The three dark dots indicate the three
superconducting phase qubits (1,2,3) embedded in the three cavities,
respectively. The interaction of qubits (1,2,3) with their cavities are
illustrated in Fig.~3(a,b,c), respectively. The interaction of the
coupler qubit $A$ with the three cavities is shown in Fig.~3(d). Since three
levels for each qubit is involved in our analysis, each qubit is renamed as a
qutrit in Fig.~3}
\label{fig:2}
\end{figure}

\section{POSSIBLE EXPERIMENTAL IMPLEMENTATION}

The physical systems composed of cavities and superconducting qubits have
been considered to be one of the most promising candidates for quantum
information processing [30-34]. In above we have considered a general type
of qubit. Let us now consider each qubit as a superconducting phase qubit
and each cavity as a one-dimensional transmission line resonator. In
addition, we assume that the coupler qubit $A$ is connected to each
resonator via a coupler capacitance. As an example of the experimental
implementation, we consider a setup in Fig. 2 for preparing the $W$ state of
three superconducting phase qubits ($1,2,3$), which are embedded in the three
one-dimensional transmission line resonators ($1,2,3$), respectively. To be more realistic, a third higher
level $\left\vert 2\right\rangle $ for each phase qubit here needs to be
considered during the operations described above, since this level $%
\left\vert 2\right\rangle $ may be excited due to the $\left\vert
1\right\rangle \leftrightarrow \left\vert 2\right\rangle $ transition
induced by the cavity mode(s), which will turn out to affect the operation
fidelity. Therefore, to quantify how well the proposed protocol works out,
we will give an analysis of the operation fidelity, by taking this higher
level $\left\vert 2\right\rangle $ into account. Because of three levels
being considered, we rename each qubit as a qutrit in the following.

When the intercavity crosstalk coupling and the unwanted $\left\vert
1\right\rangle \leftrightarrow \left\vert 2\right\rangle $ transition of
each phase qutrit are considered, the Hamiltonian (1) is modified as follows

\begin{equation}
h_I=H_I+\Theta _I,
\end{equation}
where $H_I$ is the needed interaction Hamiltonian given in Eq. (1) above,
while $\Theta _I$ is the unwanted interaction Hamiltonian, given by
\begin{eqnarray}
\Theta _I &=&\sum_{j=1}^3\widetilde{g}_j\left( e^{i\widetilde{\delta }%
_jt}a_j\sigma _{21j}^{+}+h.c.\right) +\sum_{j=1}^3\widetilde{g}_{Aj}\left(
e^{i\widetilde{\delta }_{Aj}t}a_j\sigma _{21A}^{+}+h.c.\right)  \nonumber \\
&&+\sum_{k\neq l;k,l=1}^3g_{kl}\left( e^{-i\Delta
_{kl}t}a_ka_l^{+}+h.c.\right) ,
\end{eqnarray}
where $\sigma _{21j}^{+}=\left| 2\right\rangle _j\left\langle 1\right| $ and
$\sigma _{21A}^{+}=\left| 2\right\rangle _A\left\langle 1\right| .$ The
first term represents the unwanted off-resonant coupling between the mode of
cavity $j$ and the $\left| 1\right\rangle \leftrightarrow \left|
2\right\rangle $ transition of qutrit $j$, with coupling constant $%
\widetilde{g}_j$ and detuning $\widetilde{\delta }_j=\omega _{21j}-\omega
_{cj}$ [Fig.~3(a,b,c)], while the second term indicates the unwanted
off-resonant coupling between the mode of cavity $j$ and the $\left|
1\right\rangle \leftrightarrow \left| 2\right\rangle $ transition of qutrit $%
A$, with coupling constant $\widetilde{g}_{Aj}$ and detuning $\widetilde{%
\delta }_{Aj}=\omega _{21A}-\omega _{cj}$ [Fig.~3(d)]. Here, the term
describing the cavity-induced coherent $\left| 0\right\rangle
\leftrightarrow \left| 2\right\rangle $ transition for each qutrit is not
included in the Hamiltonian $\Theta _I$, since this transition is negligible
because of $\omega _{cj}\ll \omega _{20j},\omega _{20A}$ ($j=1,2,3$)
(Fig.~3). The last term of Eq. (18) describes the intercavity crosstalk
between the three cavities, with $\Delta _{kl}=\omega _{ck}-\omega
_{cl}=\delta _l-\delta _k$ (the frequency difference between two cavities $k$
and $l$) and $g_{kl}$ (the intercavity coupling constant between two
cavities $k$ and $l$). Here and below, $kl\in \{12,13,23\}.$

\begin{figure}[tbp]
\begin{center}
\includegraphics[bb=79 283 525 769, width=13.2 cm, clip]{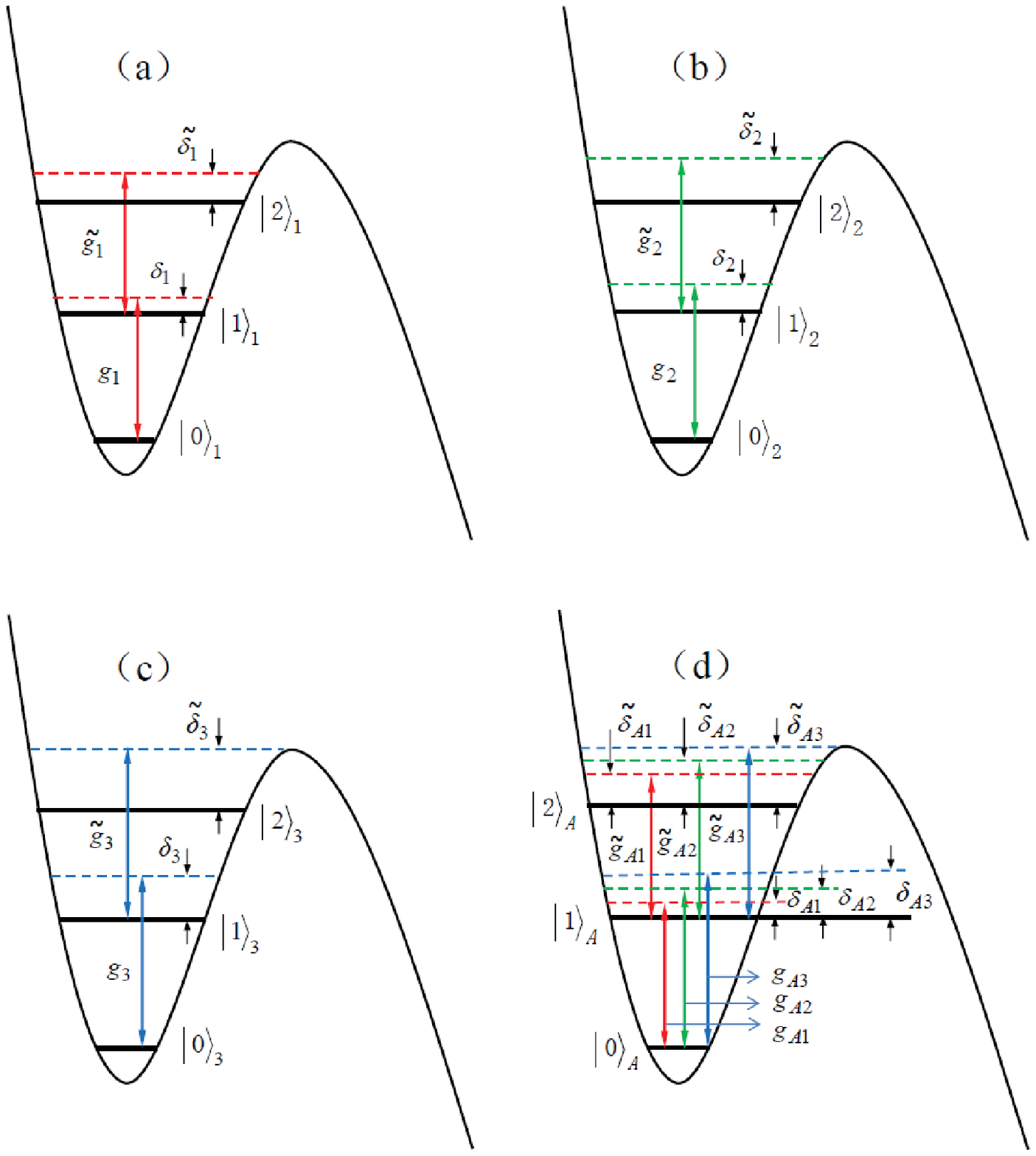} %
\vspace*{-0.08in}
\end{center}
\caption{(Color online) Illustration of qutrit-cavity interaction. (a)
Cavity $1$ is dispersively coupled to the $\left\vert 0\right\rangle
\leftrightarrow \left\vert 1\right\rangle $ transition with coupling
constant $g_{1}$ and detuning $\delta _{1},$ but far-off resonant (i.e.,
more detuned) with the $\left\vert 1\right\rangle \leftrightarrow \left\vert
2\right\rangle $ transition of qutrit $1$ with coupling consant $\widetilde{g%
}_{1}$ and detuning $\widetilde{\delta }_{1}$. (b) [and (c)] corresponds to
the case that cavity $2$ ($3$) is dispersively coupled to the $\left\vert
0\right\rangle \leftrightarrow \left\vert 1\right\rangle $ transition but
far-off resonant with the $\left\vert 1\right\rangle \leftrightarrow
\left\vert 2\right\rangle $ transition of qutrit $2$ ($3$). (d) Cavities ($%
1,2,3$) dispersively interact with the $\left\vert 0\right\rangle
\leftrightarrow \left\vert 1\right\rangle $ transition with coupling
constants ($g_{A1},g_{A2},g_{A3}$) and detunings ($\delta _{A1},\delta
_{A2},\delta _{A3}$), respectively; but they are far-off resonant with the $%
\left\vert 1\right\rangle \leftrightarrow \left\vert 2\right\rangle $
transition of qutrit $A$ with coupling constants ($\widetilde{g}_{A1},%
\widetilde{g}_{A2},\widetilde{g}_{A3}$) and detunings $\widetilde{\delta }%
_{A1},\widetilde{\delta }_{A2},\widetilde{\delta }_{A3}$), respectively.
Here, $\delta _{j}=\omega _{10j}-\omega _{cj},\widetilde{\delta }_{j}=\omega
_{21j}-\omega _{cj},\delta _{Aj}=\omega _{10A}-\omega _{cj},$ and $%
\widetilde{\delta }_{Aj}=\omega _{21A}-\omega _{cj}$ ($j=1,2,3$), where $%
\omega _{10j} $ ($\omega _{21j}$) is the $\left\vert 0\right\rangle
\leftrightarrow \left\vert 1\right\rangle $ ($\left\vert 1\right\rangle
\leftrightarrow \left\vert 2\right\rangle $) transition frequency of qutrit $%
j$, $\omega _{10A}$ ($\omega _{21A}$) is the $\left\vert 0\right\rangle
\leftrightarrow \left\vert 1\right\rangle $ ($\left\vert 1\right\rangle
\leftrightarrow \left\vert 2\right\rangle $) transition frequency of qutrit $%
A$, and $\omega _{cj}$ is the frequency of cavity $j$.}
\label{fig:3}
\end{figure}

The dynamics of the lossy system, with finite qutrit relaxation and
dephasing and photon lifetime included, is determined by the following
master equation

\begin{eqnarray}
\frac{d\rho }{dt} &=&-i\left[ h_{I},\rho \right] +\sum_{j=1}^{3}\kappa _{j}%
\mathcal{L}\left[ a_{j}\right]  \nonumber \\
&&+\sum_{j=1,2,3,A}\left\{ \gamma _{j}\mathcal{L}\left[ \sigma
_{j}^{-}\right] +\gamma _{21j}\mathcal{L}\left[ \sigma _{21j}^{-}\right]
+\gamma _{20j}\mathcal{L}\left[ \sigma _{20j}^{-}\right] \right\}  \nonumber
\\
&&+\sum_{j=1,2,3,A}\left\{ \gamma _{j,\varphi 1}\left( \sigma _{11j}\rho
\sigma _{11j}-\sigma _{11j}\rho /2-\rho \sigma _{11j}/2\right) \right\}
\nonumber \\
&&+\sum_{j=1,2,3,A}\left\{ \gamma _{j,\varphi 2}\left( \sigma _{22j}\rho
\sigma _{22j}-\sigma _{22j}\rho /2-\rho \sigma _{22j}/2\right) \right\} ,\ \
\ \
\end{eqnarray}
where $\sigma _{20j}^{-}=\left\vert 0\right\rangle _{j}\left\langle
2\right\vert ,\sigma _{20A}^{-}=\left\vert 0\right\rangle _{A}\left\langle
2\right\vert ,\sigma _{11j}=\left\vert 1\right\rangle _{j}\left\langle
1\right\vert ,\sigma _{22j}=\left\vert 2\right\rangle _{j}\left\langle
2\right\vert ;$ and $\mathcal{L}\left[ \Lambda \right] =\Lambda \rho \Lambda
^{+}-\Lambda ^{+}\Lambda \rho /2-\rho \Lambda ^{+}\Lambda /2,$ with $\Lambda
=a_{j},\sigma _{j}^{-},\sigma _{21j}^{-},\sigma _{20j}^{-}.$ Here, $\kappa
_{j}$ is the photon decay rate of cavity $a_{j}$ ($j=1,2,3$). In addition, $%
\gamma _{j}$ is the energy relaxation rate of the level $\left\vert
1\right\rangle $ of qutrit $j$, $\gamma _{21j}$ ($\gamma _{20j}$) is the
energy relaxation rate of the level $\left\vert 2\right\rangle $ of qutrit $%
j $ for the decay path $\left\vert 2\right\rangle \rightarrow \left\vert
1\right\rangle $ ($\left\vert 0\right\rangle $), and $\gamma _{j,\varphi 1}$
($\gamma _{j,\varphi 2}$) is the dephasing rate of the level $\left\vert
1\right\rangle $ ($\left\vert 2\right\rangle $) of qutrit $j$ ($j=1,2,3,A$).

The fidelity of the operation is given by
\begin{equation}
\mathcal{F}=\left\langle \psi _{\mathrm{id}}\right\vert \widetilde{\rho }%
\left\vert \psi _{\mathrm{id}}\right\rangle ,
\end{equation}
where $\left\vert \psi _{\mathrm{id}}\right\rangle $ is the output state $%
\left\vert W_{2,1}\right\rangle \left\vert 0\right\rangle _{A}\left\vert
0\right\rangle _{c1}\left\vert 0\right\rangle _{c2}\left\vert 0\right\rangle
_{c3}$ of an ideal system (i.e., without dissipation, dephasing, and
crosstalk) as discussed in the previous section; and $\widetilde{\rho }$ is
the final density operator of the system when the operation is performed in
a realistic physical \textrm{system}.

Without loss of generality, let us consider three identical superconducting
phase qutrits. According to the condition (4), we set $\delta _{1}/\left(
2\pi \right) =\delta _{A1}/\left( 2\pi \right) =-0.5$ GHz, $\delta
_{2}/\left( 2\pi \right) =\delta _{A2}/\left( 2\pi \right) =-1$ GHz, and $%
\delta _{3}/\left( 2\pi \right) =\delta _{A3}/\left( 2\pi \right) =-1.5$
GHz. For the setting here, we have $\Delta _{12}/2\pi =-0.5$ GHz, $\Delta
_{13}/2\pi =-1.0$ GHz, and $\Delta _{23}/2\pi =-0.5$ GHz. Set $\widetilde{%
\delta }_{j}=\delta _{j}-0.05\omega _{10j}$ and $\widetilde{\delta }%
_{Aj}=\delta _{Aj}-0.05\omega _{10A}$ ($j=1,2,3$)~[35]$.$ For
superconducting phase qubits, the typical qubit transition frequency is
between 4 and 10 GHz. Thus, we choose $\omega _{10A}/2\pi ,\omega
_{10j}/2\pi \sim 6.5$ GHz. Note that $g_{2}$ ($g_{3}$) is determined based
on Eq. (8), given $\delta _{1},$ $\delta _{2}$ ($\delta _{3}$)$,$ and $%
g_{1}. $ In addition, $g_{Aj}$ is determined by Eq. (15)$,$ given $g_{j}$ ($%
j=1,2,3$). For the present case, we have $n=3.$ Next, one has $\widetilde{g}%
_{j}\sim \sqrt{2}g_{j}$ and $\widetilde{g}_{Aj}\sim \sqrt{2}g_{Aj}$ ($%
j=1,2,3 $) for the phase qutrit here. We choose $\gamma _{j,\varphi
1}^{-1}=\gamma _{j,\varphi 2}^{-1}=2.5$ $\mu $s, $\gamma _{j}^{-1}=10$ $\mu $%
s, $\gamma _{21j}^{-1}=7.5$ $\mu $s, and $\gamma _{20j}^{-1}=30$ $\mu $s;
and $\kappa _{j}^{-1}=5$ $\mu $s ($j=1,2,3$). For a phase qutrit with the
three levels considered here, the $\left\vert 0\right\rangle \leftrightarrow
\left\vert 2\right\rangle $ dipole matrix element is much smaller than that
of the $\left\vert 0\right\rangle \leftrightarrow \left\vert 1\right\rangle $
and $\left\vert 1\right\rangle \leftrightarrow \left\vert 2\right\rangle $
transitions. Thus, $\gamma _{20j}^{-1}\gg \gamma _{10j}^{-1},\gamma
_{21j}^{-1}.$

\begin{figure}[tbp]
\begin{center}
\includegraphics[bb=0 0 800 500, width=10.5 cm, clip]{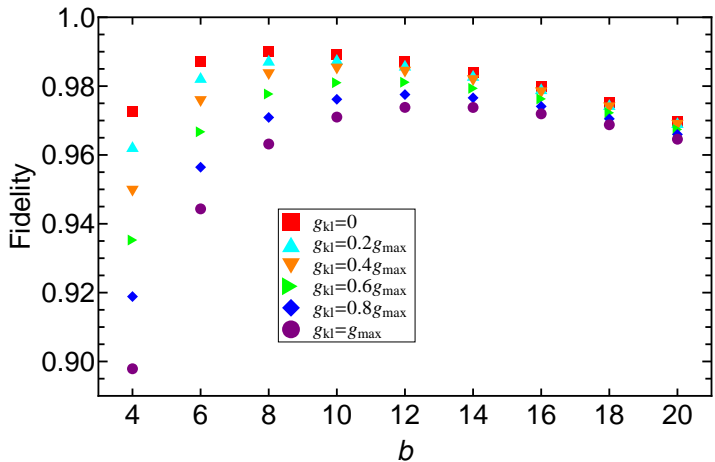} %
\vspace*{-0.08in}
\end{center}
\caption{(Color online) Fidelity of the $W$-state preparation versus the
normalized detuning $b=\left\vert \delta _{1}\right\vert /g_{1}$. Refer to
the text for the parameters used in the numerical calculation. Here, $g_{kl}$
are the coupling strengths between cavities $k$ and $l$ ( $k\neq l;\,k,l\in
\left\{ 1,2,3\right\} $), which are taken to be the same for simplicity.}
\label{fig:4}
\end{figure}

For the parameters chosen above, the fidelity versus $b=\left\vert \delta
_{1}\right\vert /g_{1}$ is plotted in Fig.~4 for $g_{kl}=0,0.2g_{\max },$ $%
0.4g_{\max },0.6g_{\max },0.8g_{\max },g_{\max },$ where $g_{\max }=\max
\{g_{A1},g_{A2},g_{A3}\}.$ Fig. 4 shows that for $g_{kl}\leq 0.2g_{\max }$,
the effect of intercavity cross coupling between the three cavities on the
operational fidelity is negligible, which can be seen by comparing the
top two curves. In addition, it can be seen from Fig.~4 that for $b\sim 8$
and $g_{kl}=0.2g_{\max },$ a high fidelity $\sim 99\%$ is available for the $%
W$-state preparation.

The condition $g_{kl}\leq 0.2g_{\max }$ is not difficult to satisfy with the
typical capacitive cavity-qutrit coupling illustrated in Fig.~2. As long as
the cavities are physically well separated, the intercavity cross-talk
coupling strength is $g_{kl}\sim g_{Ak}C_{l}/C_{\Sigma
},g_{Al}C_{k}/C_{\Sigma },$ where\ $C_{\Sigma }=C_{1}+C_{2}+C_{3}+C_{q}$.
With a choice of $C_{1},C_{2},C_{3}\sim 1$ fF and $C_{\Sigma }\sim 10^{2}$
fF (the typical values of the cavity-qutrit coupling capacitance and the sum
of all coupling capacitance and qutrit self-capacitance, respectively), one
has $g_{kl}\sim 0.01g_{Ak},0.01g_{Al}$. Because of $g_{A1,}$ $g_{A2,}$ $%
g_{A3}\leq g_{\max },$ the condition $g_{kl}\leq 0.2g_{\max }$ can be
readily met in experiments. Thus, it is straightforward to implement designs
with sufficiently weak direct intercavity couplings.

For $b\sim 8$, we have $\{g_{1},g_{2},g_{3},g_{A1},g_{A2},g_{A3}\}\sim $ $%
\{62.5,88.4,108.3,36.1,51.0,62.5\}$ MHz. Experimentally, a coupling constant
$\sim 220$ MHz can be reached for a superconducting qutrit coupled to a
one-dimensional CPW (coplanar waveguide) resonator [36,37], and that $T_{1}$
and $T_{2}$ can be made to be on the order of $10-100$ $\mu $s or longer for
state-of-the-art superconducting devices [38-42]. For phase \textrm{qutrits,
t}he energy relaxation time $T_{1}^{^{\prime }}$ and dephasing time $%
T_{2}^{\prime }$ of the level $\left\vert 2\right\rangle $ are,
respectively, comparable to $T_{1}$ and $T_{2}$ because of $T_{1}^{\prime
}\sim T_{1}/\sqrt{2}$ and $T_{2}^{^{\prime }}\sim T_{2}.$ With $\omega
_{10A}/2\pi ,\omega _{10j}/2\pi \sim 6.5$ GHz chosen above, we have $\omega
_{c1}/2\pi \sim 6.0$ GHz, $\omega _{c2}/2\pi \sim $ $5.5$ GHz, and $\omega
_{c3}/2\pi \sim $ $5.0$ GHZ. For these cavity frequencies and the values of $%
\kappa _{1}^{-1},\kappa _{2}^{-1}$ and $\kappa _{3}^{-1}$ used in the
numerical calculation, the required quality factors for the three cavities
are $Q_{1}\sim 1.9\times 10^{5},$ $Q_{2}\sim 1.7\times 10^{5},$ and $%
Q_{3}\sim 1.6\times 10^{5},$ respectively. It should be mentioned that
superconducting CPW resonators with a loaded quality factor $Q\sim 10^{6}$
have been experimentally demonstrated [43,44], and planar superconducting
resonators with internal quality factors above one million ($Q>10^{6}$) have
also been recently reported [45]. Our analysis given here demonstrates that
high-fidelity preparation of the $W$ state of three intracavity qubits by
using this proposal is feasible within the present circuit QED technique. We
remark that further investigation is needed for each particular experimental
setup. However, it requires a rather lengthy and complex analysis, which is
beyond the scope of this theoretical work.

\section{CONCLUSION}

We have proposed a general method to generate the $W$-class entangled states of $n$
qubits distributed in different $n$ cavities. As shown above, this proposal
offers some advantages and features: the entanglement preparation does not
employ cavity photons as quantum buses, thus decoherence caused due to the
cavity decay is greatly suppressed during the operation; only one coupler
qubit is needed to connect with all cavities such that the circuit complex
is greatly reduced; moreover, only one step of operation is required and no
classical pulse is needed, so that the operation is much simplified. The time required decreases
as the number of qubits increases. In addition, our numerical simulation shows that high-fidelity implementation
of the three-qubit $W$ state is feasible for the current circuit QED
technology. The method presented here is also applicable to a wide range of physical
implementations with different types of qubits such as quantum
dots, superconducting qubits (e.g., phase, flux and charge qubits), NV
centers, and atoms.


\begin{thebibliography}{99}
\bibitem{35}  For a phase qutrit, a ratio $5\%$ of the anharmonicity between
the $\left\vert 0\right\rangle \leftrightarrow \left\vert 1\right\rangle $
transition frequency and the $\left\vert 1\right\rangle \leftrightarrow
\left\vert 2\right\rangle $ transition frequency to the the $\left\vert
1\right\rangle \leftrightarrow \left\vert 2\right\rangle $ transition
frequency is readily achieved in experiments.
\end{thebibliography}

\section*{ACKNOWLEDGMENTS}

X.L.H. acknowledges the funding support from the Zhejiang Natural Science Foundation under Grant
No. LY12A04008. F.Y.Z. acknowledges the funding support from the National Science Foundation of China under Grant No. 11175033.
C.P.Y. was supported in part by the National Natural Science Foundation of China under Grant Nos.
11074062 and 11374083, the Zhejiang Natural Science Foundation under Grant
No. LZ13A040002, and the funds from Hangzhou Normal University under Grant
Nos. HSQK0081 and PD13002004. This work was also supported by the funds from Hangzhou City
for the Hangzhou-City Quantum Information and Quantum Optics Innovation Research Team.

\end{document}